# Superconductor-insulator transition in fcc-$GeSb_2Te_4$ at elevated pressures

Bar Hen [1*], Samar Layek [1], Moshe Goldstein [1], Victor Shelukhin [1], Mark Shulman [1], Michael Karpovski [1], Eran Greenberg [2], Eran Sterer [3], Yoram Dagan [1], Gregory Kh. Rozenberg [1], Alexander Palevski [1]

[1] *Raymond and Beverly Sackler School of Physics and Astronomy, Tel-Aviv University, Tel Aviv 69978, Israel*
[2] *Center for Advanced Radiation Sources, University of Chicago, Chicago, Illinois 60637, USA*
[3] *Physics Department, Nuclear Research Center Negev, P.O. Box 9001, Beer-Sheva 84190,Israel*

We show that polycrystalline $GeSb_2Te_4$ in the *fcc* phase (*f*-GST), which is an insulator at low temperature at ambient pressure, becomes a superconductor at elevated pressures. Our study of the superconductor –insulator transition versus pressure at low temperatures reveals a second order quantum phase transition with linear scaling (critical exponent close to unity) of the transition temperature with the pressure above the critical zero-temperature pressure. In addition, we demonstrate that at higher pressures the *f*-GST goes through a structural phase transition via amorphization to *bcc* GST (*b*-GST), which also become superconducting. We also find that the pressure regime where an inhomogeneous mixture of amorphous and *b*-GST exists, there is an anomalous peak in magnetoresistance, and suggest an explanation for this anomaly.

*PACS numbers: 74.10.+v; 74.62.Fj; 61.50Ks; 73.43.Nq*

## Introduction

$GeSb_2Te_4$ (GST) is a phase-change-material, whose unusual physical properties [1, 2, 3, 4, 5, 6] promise many potential applications in the electronics industry [7, 8, 9, 10].
One of the newly discovered properties of the GST is the emergence of the superconductivity under elevated pressures [11]. In our previous high-pressure study of GST [11], superconductivity was observed in amorphous GST (*a*-GST), orthorhombic GST (*o*-GST) and in *bcc* GST (*b*-GST). In addition we have demonstrated [11] that hexagonal GST remained in the normal state for the entire range of available temperatures and pressures. However, the transport properties of *fcc* GST (*f*-GST) at elevated pressure and at low temperatures remained unexplored.

This paper is devoted to the study of the properties of GST material in the *fcc* phase at high pressure and low temperatures. We demonstrate that *f*-GST undergoes a superconductor to insulator transition (SIT) at low temperatures when the pressure is applied as an external control parameter. We find that the superconducting transition temperature vanishes linearly with pressure, while the GST remains in the *f*-GST phase, strongly suggesting a second-order quantum phase transition (QPT) with a critical exponent close to unity.
The observed appearance of superconductivity is preceded by a significant change in the normal state resistance of the samples by a few orders of magnitude. Furthermore, we demonstrate that superconductivity with somewhat higher $T_c$ appears at higher pressures, when *b*-GST starts to form. In the region where these two phases coexist, an anomalous behavior of the magnetoresistance is observed, whereby a sharp resistance peak appears in the vicinity of the upper critical field. We suggest an explanation for this behavior.

## Experimental

In our transport and XRD experiments, we have used the following procedure for the preparation of *f*-GST samples. Initially, the few micron thick GST films were sputtered from a commercial target of *h*-

GST (hexagonal $GeSb_2Te_4$). As we reported earlier [11], the films sputtered onto a room temperature substrate are amorphous (*a*-GST). An atomic composition and morphology of the as-prepared *a*-GST film was checked by scanning electron microscopy (SEM), energy dispersive X-ray spectroscopy (EDS) and X-ray photoelectron spectroscopy (XPS) analysis [11]. The annealing of the sputtered films at 146 ℃ causes the transformation of the *a*-GST into an *fcc* polycrystalline phase. An X-ray diffraction (XRD) analysis confirming the formation of the *fcc* phase of the annealed films is shown in Fig. 1(a) at 0 GPa. Finally, a powder of *f*-GST was prepared by the mechanical removal of the *f*-GST film from the substrate.

Pressure was exerted using miniature diamond anvil cells (DACs) [12] with diamond anvil culets of 250 μm. A pre-indented stainless-steel or rhenium gasket was drilled and then filled and covered with a powder layer of 75% $Al_2O_3$ and 25% NaCl for electrical insulation. The powder of *f*-GST was placed onto the culets. A Pt foil with a thickness of 5-7 μm was cut into triangular probes connecting between the sample and copper leads allowing the electrical transport measurements at elevated pressures. In each DAC 6-8 probes were placed. Fig. 2(a) depicts a setup of 6 Pt foils (bright areas) between the diamond and the sample (dark areas) in a four-probe configuration. Ruby was used as a pressure gauge.

Electrical transport measurements were performed using a $^4$He cryostat. The sample was compressed up to 44 GPa in increments of 2 GPa on average, and cooled down from ambient temperature down to 1.4 K. After each pressure increment a temperature cycle was performed.

Synchrotron XRD measurements of *f*-GST powder were performed at room temperature up to 47 GPa at the beamlines 13ID-D and 13-BM-C of APS (Argonne, IL, USA), with wavelengths of λ = 0.3738 and 0.434 Å, respectively, in angle-dispersive mode with patterns collected using a MAR CCD detector. The image data were integrated using DIOPTAS [13] and the resulting diffraction patterns were analyzed with the GSAS+EXPGUI [14, 15] program. XRD data at ambient temperature and pressure have been collected in symmetric Bragg-Brentano geometry with $CuK_\alpha$ radiation (λ = 1.5406 Å) on Bruker D8 Discover Θ:Θ X-ray diffractometer equipped with one-dimensional LynxEye XE detector.

## Experimental results of XRD and transport studies

We start the description of our experimental results with XRD and transport studies at room temperature.

As depicted in Fig. 1(a) and (b) the diffraction patterns obtained from both slots of measurements show the existence of the *fcc* phase up to about 14 GPa. The broadening and shifting of the (200) and (400) peaks along with the disappearance of the other *fcc* peaks of GST is clearly observed for pressures above 14 GPa, indicating the amorphization of the *fcc* phase. Upon further increase in pressure the emergence of the *bcc* phase of GST becomes evident at pressures exceeding 29 GPa Fig. 1(a). The observed amorphization as well as the formation of the *bcc* phase are consistent with previously reported results [16]. We would like to note that the appearance of the intermediate orthorhombic phase reported in [16] was not detected in our data. The change in density versus pressure which is depicted in Fig. 2(c) is fitted to the second order Birch-Murnaghan (BM2) equation of state (EOS) [17] with the extracted parameters indicated in the figure labels.

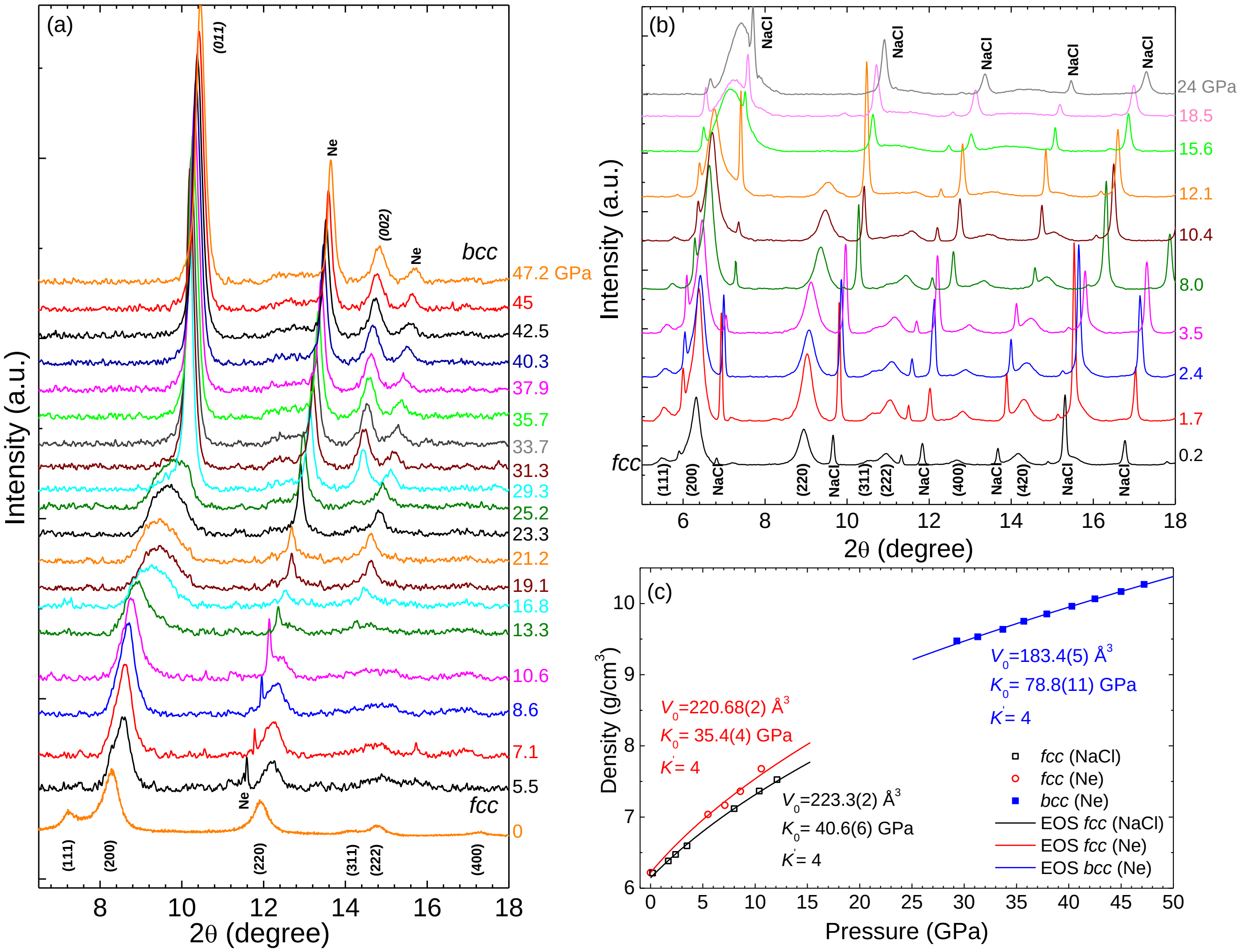

*FIG 1. X-ray diffraction upon compression as a function of pressure at T=298 K for two slots of measurements at 2 different beam lines and the density of the observed phases. Miller indices indicate the observed reflections of the fcc and bcc phases. (a) At the range of pressure from 5.5 GPa to 47.2 GPa with Ne as pressure medium and $\lambda$ =0.434 Å (0 GPa data was taken from the annealed film on the substrate). (b) At the range of pressure from 0.2 GPa to 24 GPa with NaCl as pressure medium and $\lambda$=0.3344 Å. (c) Calculated density as a function of pressure and the comparison with the theoretical "BM2 curve".* $K_0$, and $V_0$ are the bulk modulus, and the volume per formula at 1 bar and 300 K, respectively. *The vertical error bars do not exceed the size of the symbols.*

As depicted in Fig. 2(b), the room temperature resistance of *f*-GST drops very sharply (by more than 2 orders of magnitude) as a result of the application of just a few GPa. This sharp decrease is followed by a more moderate drop of one order of magnitude as a result of compression of the sample to about 8 GPa. For pressures above 8 GPa, the resistance remains roughly constant. We would like to emphasize that the resistance drop is not accompanied by any crystallographic change as already mentioned above (Fig. 1). The observed slight increase in resistance (by a factor of 2) corresponds to the pressure range where the amorphization is observed, namely coinciding with the region between *f*-GST and *b*-GST. For pressures above 25 GPa the value of the resistance remains roughly constant.

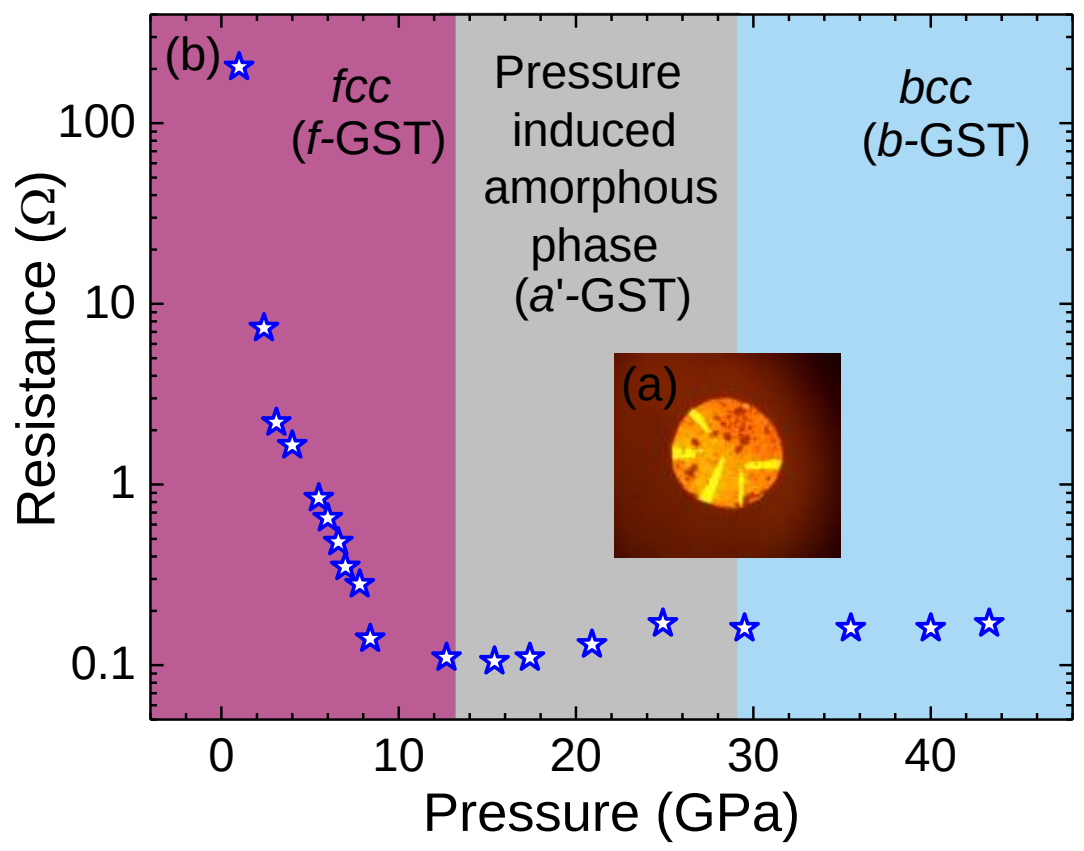

*FIG 2. Room temperature resistance versus pressure. (a) Contact configuration. (b) Resistance as a function of pressure. The colored regions are marked according to our XRD data from Fig. 1.*

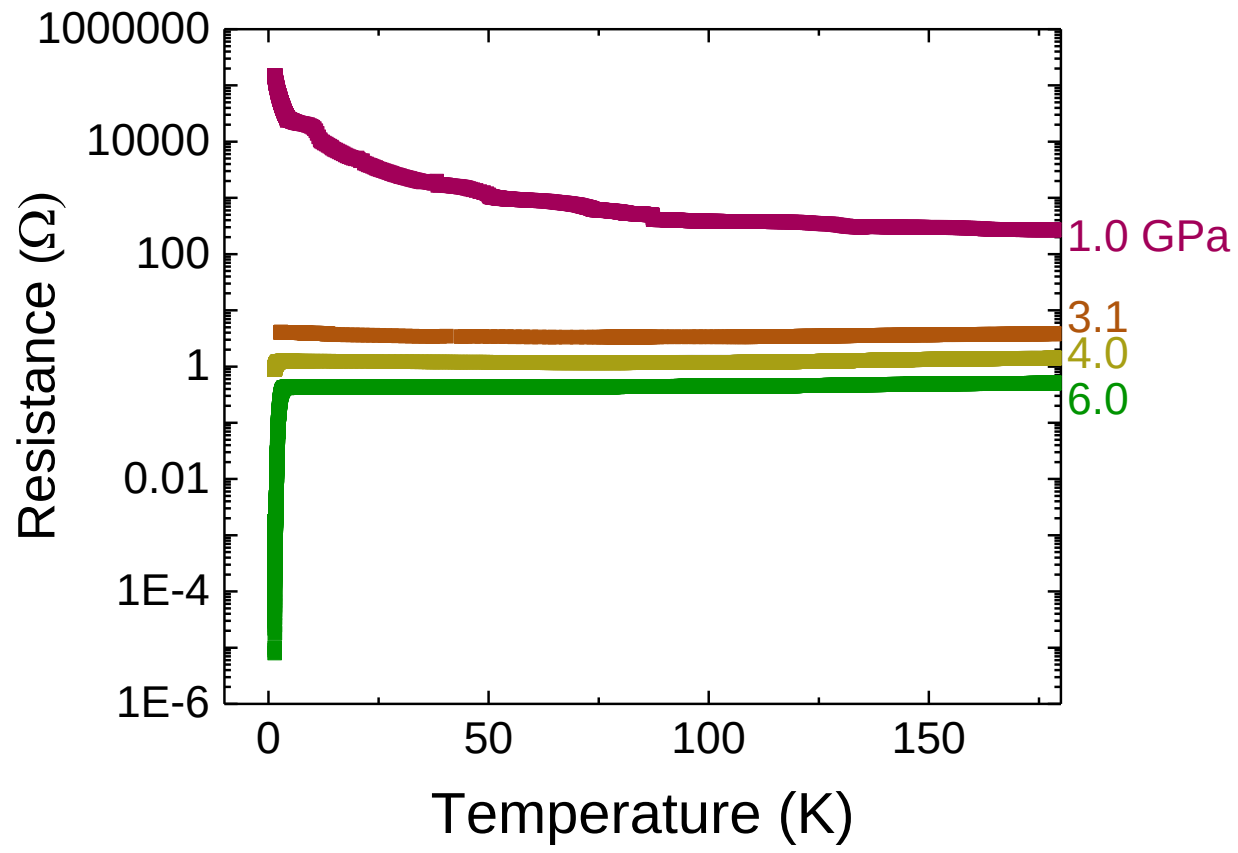

*FIG 3. Temperature dependence of the first pressure points demonstrating the SIT in f-GST.*

We now present the results of our transport measurements at low temperatures. The resistance versus temperature at pressures between 1-6 GPa reveals a clear SIT type behavior in *f*-GST, as depicted in Fig. 3. One can clearly see the large resistance change during the transition from an insulating state at 1.0 GPa to the full superconducting state at 6.0 GPa. It is also evident that the onset of superconductivity appears at 4.0 GPa.

Fig. 4(a) focuses on the superconductivity transitions in the pressure range where the samples remain in the *fcc* phase (pressures up to 12.7 GPa, cf. Fig. 2(b) in purple region). Throughout the paper, the definition for critical temperature is that of the temperature at which the resistance equals half of the normal state resistance immediately above the transition. The superconducting critical temperature $T_c$ increases roughly linearly from 1.8 K at 4.0 GPa to 5.8 K at 10.4 GPa. The linear dependence is emphasized by the straight dashed trend line in Fig. 4(b), which extrapolates to zero temperature at $P_{c,0}$=3.1 GPa. As will be discussed below, this linear dependence is expected from a Ginzburg-Landau type mean field theory.

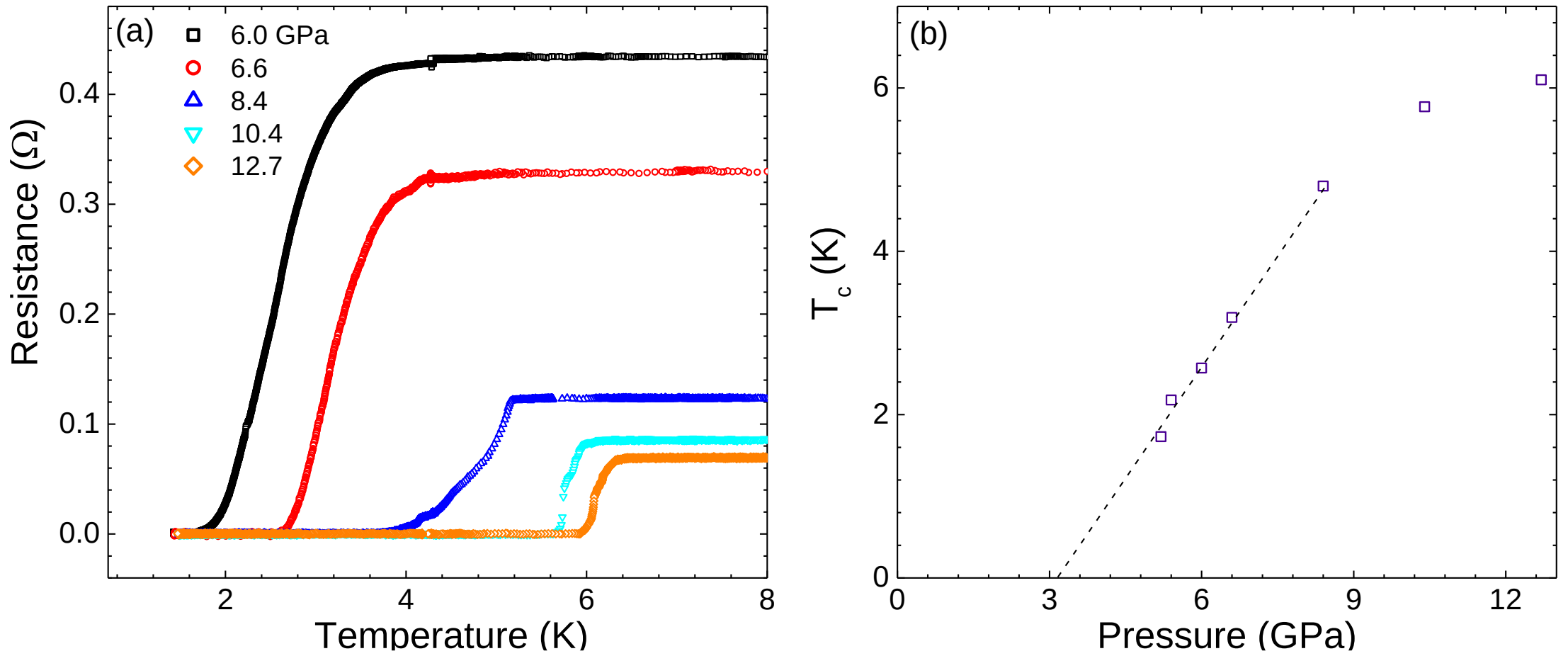

*FIG 4. Superconductivity transitions of f-GST. (a) Resistance as a function of temperature at various pressures showing complete transitions (resistance drops to zero). (b) Critical temperature as a function of pressure. The first pressure points follow a linear trend (dashed line).* The vertical error bars do not exceed the size of the symbols.

Upon further increase of the pressure, in the range between 12 GPa and 27 GPa, the superconducting transition temperature shows saturation, as can be inferred from Fig. 5(a). In this pressure range, the pressure induced amorphous phase forms (Fig. 2) and one may expect the coexistence of *f*-GST and the amorphous phase due to some pressure inhomogeneity inside the pressure cell. The fact that there is only a single transition in the curves for this pressure range, along with the moderate increase in $T_c$ up to 27.0 GPa, suggest that the pressure induced amorphous phase is similar to *f*-GST, at least in its superconductivity properties. This assumption is in good agreement with recent theoretical simulations [18], which show a formation of an amorphous structure of cubic framework for GST at pressures above 18 GPa (hereafter referred to as *a'*-GST), characterized by the collapse of long range order, formation of homopolar bonds and slight increase of coordination numbers. Furthermore, according to [19], at ~27 GPa strong distortions in the crystal structure are observed resulting in formation of *bcc* phase. These results correspond well with our experimental observations.

For pressures exceeding 27.0 GPa, two distinct transitions appear in the resistance vs. *T* curves, as shown in Fig. 5(b). This signifies the appearance of the *b*-GST phase, in accordance with the XRD data (Fig. 2(b) in blue region). These double transitions can be interpreted as coexistence of *a*'-GST with *b*-GST, both being superconductors at different temperatures. The observed coexistence of both phases throughout a wide range of pressures is most probably due to inhomogeneous pressure distribution inside the cell ($Al_2O_3$+NaCl is considered a poor pressure medium relative to Ne which is used for XRD measurements). We associate the higher $T_c$ value with *b*-GST, since it is apparent that the critical temperature of *a*'-GST has already been saturated at about 6.6 K and the higher value for *b*-GST is consistent with our previously reported results for this phase [11].

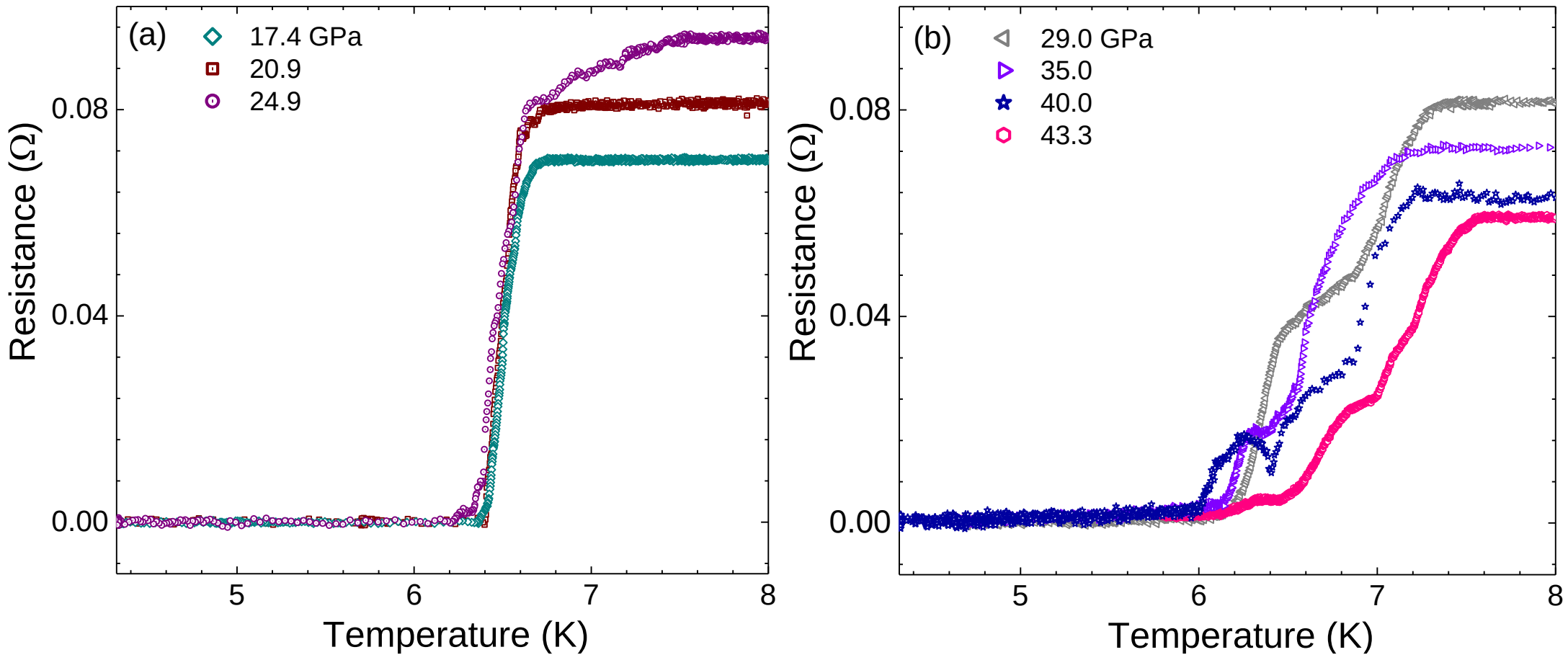

*FIG 5. Superconductivity transitions of (a) a'-GST and (b) a'-GST and b-GST mixture. The double transitions observed in (b) are interpreted as a mixture of a'-GST and b-GST.*

A summary of the critical temperature dependence on pressure results in the *T-P* phase diagram shown in Fig. 6. The two distinct critical temperatures are deduced from the analysis of Fig. 5(b), where the $T_c$ for each phase is defined by the mid-value of the corresponding resistance drop.

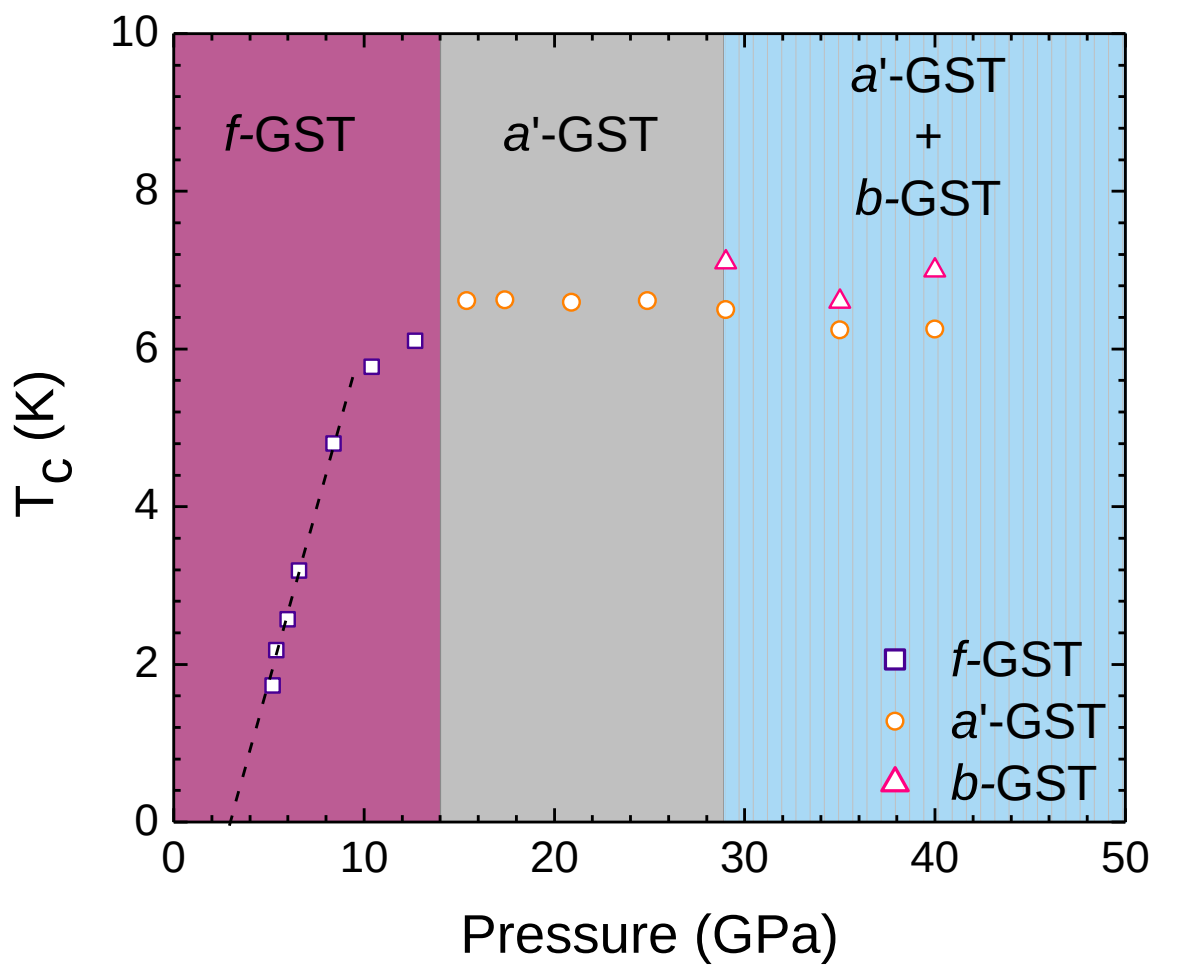

*FIG 6. Superconducting phase diagram. The colored regions are marked according to our XRD data (Figs. 1 and 2(b)) and our stupeconductivity results (Figs. 4(a) and 5).* The vertical error bars do not exceed the size of the symbols.

The appearance of the double transitions is accompanied by the observation of anomalous magnetoresistance at *T*= 4.2 K at different pressures, as shown in Fig. 7. Fig. 7(a) reveals that for pressures below 29.0 GPa the magnetoresistance behaves as expected – a distinct normal transition from the superconducting state to the normal state for all pressures for which $T_c$>4.2 K, with a well-defined upper critical field $H_{c2}$. However, at higher pressures where a considerable fraction of the sample transforms into *b*-GST, we observe an anomalous behavior, where the resistance sharply increases above the normal state resistance, followed by a drop to its normal value (Fig. 7(b)). This peak starts appearing at 35.0 GPa, becomes most-pronounced at 36.0 GPa (where the peak reaches 1.5 times the value of the normal state resistance), and then gradually decreases, practically disappearing at 43.3 GPa, where the entire sample is probably in a single *b*-GST phase.

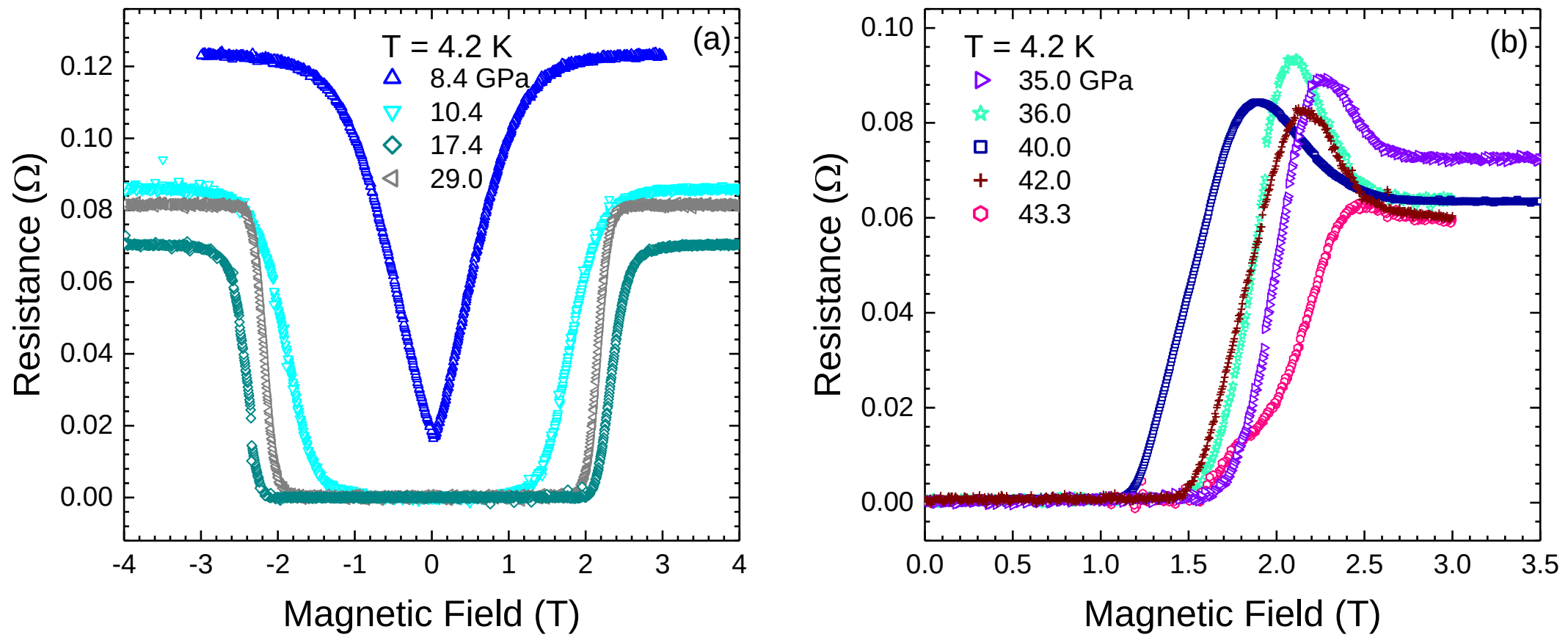

*FIG 7. Magnetoresistance at 4.2 K for (a) pressures up to 29.0 GPa and (b) pressures above 29.0 GPa. In (a) normal transitions are observed, whereas the curves in (b) exhibit an anomalous behavior, with the resistance sharply increasing above its normal state value.*

**Discussion and Analysis of the results**

We now turn to the analysis of our experimental findings. Let us start with the linear increase of $T_c$ for pressures immediately above the zero-temperature critical pressure, $P_{c,0} = 3.1$ GPa (Fig. 4(b)). In Ginzburg-Landau theory [20], in the absence of a magnetic field, the superconducting part of the free energy density can be expanded near the transition to 4th order in the order parameter $\psi$,

$$f_{SC} = \alpha(T,P)|\psi|^2 + \beta(T,P)|\psi|^4,$$

where the coefficients $\alpha$ and $\beta$ are now not only functions of the temperature $T$, but also of the pressure $P$. As usual, $\beta > 0$ to ensure the finiteness of $|\psi|$ at the minimum, hence its exact $T$ and $P$ dependence is irrelevant near the transition. As for $\alpha$, it is positive in the normal phase, and negative in the superconducting phase. Since it vanishes at the transition, in its vicinity it can be expanded to linear order in temperature and pressure,

$$\alpha(T,P) \sim AT + BP + C = A\big(T - T_C(P)\big),$$

where $A > 0$ (as usual), and furthermore, $T_c(P) = -(B/A)(P - P_{c,0})$ with $P_{c,0} = -C/B$ (and hence $B < 0, C > 0$). Thus, in Ginzburg-Landau Theory $T_c$ should indeed be a linear function of $P$ close to the zero-temperature critical pressure $P_{c,0}$. While this is a mean-field prediction, Ginzburg-Landau theory is known to give a good quantitative description of the superconducting transition in 3D, due to the typical extreme smallness of the Ginzburg number. Moreover, in the vicinity of the quantum critical point at $T = 0, P = P_{c,0}$, the system is effectively 4-dimensional (counting also the time axis), and mean-field theory becomes an even better approximation [21].

Let us now turn to the anomalous peak in magnetoresistance (Fig. 7(b)). It occurs at the pressure range where *b*-GST starts to form. In this range, as we already mentioned, we observe a double transition as a function of temperature at zero field, indicating the coexistence of the *a*'-GST and *b*-GST phases. It is reasonable to associate the appearance of the anomalous magnetoresistance with the formation of an inhomogeneous mixture of these two phases. One possible scenario for such an anomalous peak in the magnetoresistance has been discussed in theoretical papers [22, 23, 24] trying to explain the huge magnetoresistance peak observed in superconductor thin films of InO [25, 26] and TiN [27, 28]. In these models, the experimental system is viewed as a 2D array of Josephson-coupled superconducting islands at zero magnetic field. It is argued that such a system possesses a highly resistive state when the magnetic field is large enough to suppress the coherence between the islands, while not being large enough to destroy the superconductivity in each island. Although it is possible that the anomalous MR observed in our 3D system has a similar origin, there is an alternative explanation which might be more relevant to our system.

In a system where two structural phases coexist, there should exist a range of magnetic fields where one phase is superconducting while the other is normal. The finite superconducting gap suppresses the transmission of quasi-particles between the superconducting and normal regions at low temperature. On the other hand, Andreev reflections are still allowed. In this process Cooper pairs are transmitted into the superconductor while the electrons are reflected as holes into the normal phase. However, when the transparency of the interface between the phases is low, the tunneling probability of pairs is strongly suppressed [29]. This implies that the resistance of a percolating phase in the normal with non-percolating islands of a different phase might be larger when these islands are superconducting (intermediate magnetic fields) than when the islands are normal (high magnetic fields).

Now, in our samples we observe two superconducting transitions as a function of temperature at zero field and in the pressure range of 29-40 GPa (Fig. 5(b)), it is reasonable to assume that the *b*-GST, which has a higher transition temperature, and thus presumably also a higher critical field, is not percolating between the contacts, since otherwise there would be only one transition, when *b*-GST becomes superconducting. Therefore, the anomalous MR observed in our samples for some pressure values in the above-mentioned range can be explained as follows. For low magnetic fields, both percolating *a*'-GST regions of the sample and isolated islands of *b*-GST are in the superconducting state, resulting in the zero resistance of the sample. When the upper critical field of *a*'-GST is approached, the resistance starts to rise and reaches the values above the normal state resistance, since the *b*-GST remains in the superconducting state. The sample resistance starts to decrease towards the normal state resistance only after the superconductivity is destroyed in the *b*-GST islands, namely when the upper critical field of *b*-GST is reached.

## Conclusions

To summarize, we demonstrated that the polycrystalline $GeSb_2Te_4$ in *fcc* phase becomes a superconductor at elevated pressure. The linear variation of the superconductor transition temperature versus pressure indicates a second-order quantum phase transition. The linear extrapolation to zero temperature gives the value of the quantum critical point – the critical pressure of $P_{c,0} = 3.1$ GPa. In addition, we demonstrate that at higher pressures the *f*-GST goes through structural phase transition via amorphization to *b*-GST, with all phases exhibiting superconductivity. We also provided a possible explanation for the peak in magnetoresistance observed in the pressure range where inhomogeneous mixture of *a*'-GST and *b*-GST is present.

## Acknowledgements:

The authors would like to thank Aharon Kapitulnik, Leonid Glazman and Boris Spivak for very helpful discussions. We are grateful to Yishay Feldman for the XRD measurement at ambient pressure and fruitful discussions.

The authors thank Dongzhou Zhang and Vitali Prakapenka for their assistance with the synchrotron XRD measurements. Portions of this work were performed at GeoSoilEnviroCARS (The University of Chicago, Sector 13), Advanced Photon Source (APS), Argonne National Laboratory. GeoSoilEnviroCARS is supported by the National Science Foundation - Earth Sciences (EAR - 1634415) and Department of Energy-GeoSciences (DE-FG02-94ER14466). This research used resources of the Advanced Photon Source, a U.S. Department of Energy (DOE) Office of Science User Facility operated for the DOE Office of Science by Argonne National Laboratory under Contract No. DE-AC02-06CH11357. This research was partially funded by PAZI foundation under grant number 268/15. Support from the Israeli Science Foundation (grant numbers 277/16, 1189/14, and 227/15), the Israel Ministry of Science and Technology under (contracts number 3-11875 and 3-12419), the German-Israeli Science Foundation (grant I-1259-303.10/2014), and the US-Israel Binational Science Foundation (grants 2014262 and 2014098), is gratefully acknowledged.

* barh@post.tau.ac.il